\documentclass[twocolumn,showpacs,preprintnumbers,amsmath,amssymb]{revtex4}

\usepackage{graphicx}
\usepackage{dcolumn}
\usepackage{bm}

\begin{document}

\title{Pseudogap Behavior Revealed in Interlayer Tunneling in Overdoped Bi$_2$Sr$_2$CaCu$_2$O$_{8+x}$}

\author{Myung-Ho Bae$^{1,2}$}
\author{Jae-Hyun Park$^1$}
\author{Jae-Hyun Choi$^1$}
\author{Hu-Jong Lee$^{1,3}$}
\author{Kee-Su Park$^4$}

\affiliation{$^1$Department of Physics, Pohang University of Science
and Technology, Pohang 790-784, Republic of Korea}%
\affiliation{$^2$Department of Physics, University of Illinois at
Urbana-Champaign, Urbana, Illinois 61801-3080, USA}
\affiliation{$^3$National Center for Nanomaterials Technology,
Pohang 790-784, Republic of Korea}
\affiliation{$^4$Department of
Physics, Pusan National University, Busan 609-735, Republic of
Korea}

\date{\today}

\begin{abstract}
We report heating-compensated interlayer tunneling spectroscopy
(ITS) performed on stacks of overdoped
Bi$_2$Sr$_2$CaCu$_2$O$_{8+x}$ intrinsic junctions, where most of
bias-induced heating in the ITS was eliminated. The onset
temperature of the pseudogap (PG), revealed in the hump structure
of the electronic excitation spectra, reached nearly room
temperature for our overdoped intrinsic junctions, which
represented the genuine PG onset. At a temperature below but close
to $T_c$, both the superconducting coherence peak and the
pseudogap hump coexisted, implying that the two gaps are of
separate origins. The hump voltage increased below $T_c$,
following the superconducting gap voltage, which led to a
conclusion that the hump structure below $T_c$ in our ITS arose
from the combined contribution of the quasiparticle spectral
weights of two different characters; one of the superconducting
state and another of the PG state near the antinodal region.

\end{abstract}

\maketitle

\section{Introduction}

Conventional superconductors in their superconducting state are
characterized by opening of the superconducting gap (SG) in the
electronic density of states (DOS). Superconductivity appears when
electrons bind into Cooper pairs and condense with long-range
order below the superconducting transition temperature $T_c$.
Cuprate superconductors, however, as one of the most intriguing
characteristics in their normal state, show the unusual emergence
of the pseudogap (PG) in the electronic excitation spectrum even
above $T_c$, which persists up to a temperature $T^*$, the PG
onset temperature. It has been widely accepted that understanding
the PG origin and the relation between the PG and the SG may lead
to a key to finding the basic mechanism of high-temperature
superconductivity,\cite{Timusk} which is not fully resolved up to
the present.

There are two schools of thought as to understanding the PG in the
cuprate physics: one-gap and two-gap ones. One-gap school regards
the PG as the precursor of the SG, where thermal fluctuations
destroy long-range order while maintaining gap-like features in
the excitation spectra in a certain high-temperature range ($T \ge
T_c$) of the normal state. Thus, the PG in question is believed to
bring about the partial depletion of the DOS at the normal-state
Fermi surface\cite{Damascelli}, resulting in the Fermi
arcs.\cite{Arc} The other school interprets the PG, especially in
the underdoped regime, in terms of two gaps; a small SG revealed
in the nodal regions and a large gap of different origin in the
antinodal regions. In the two-gap model, the Fermi arcs are
believed to emerge due to long-range order that, however, is not
associated with the superconducting order. Above $T_c$ the SG may
vanish, leaving the other long-range order connected to the Fermi
arcs, Fermi surface nestings\cite{Dessau,Gofron, Gabovich} or
Fermi surface pockets\cite{Varma}. Recent Raman and
angle-resolved-photoemission-spectroscopy (ARPES) measurements
show the consequences of opening of two gaps in underdoped
single-layer HgBa$_2$CuO$_{4+\delta}$ superconductors and bilayer
Bi$_2$Sr$_2$Ca$_{1-x}$Y$_x$Cu$_2$O$_8$ superconductors,
respectively.\cite{Tacon, Tanaka}

The surface tunneling studies on the PG behavior suggest that the PG
can evolve into the SG in the norm of one gap.\cite{Fischer} In
interlayer tunneling measurements on densely stacked intrinsic
Josephson junctions (IJJs)\cite{Kleiner} formed in the layered
cuprates, however, it has been proposed that the SG vanishes at
$T_c$ and the PG may exist both below and above $T_c$. This
experimental observation from IJJs is claimed to provide the norm of
two gaps, where the SG and the PG are considered to be of different
origins.\cite{ITS} The interlayer tunneling reveals the intrinsic
bulk tunneling properties between CuO$_2$ superconducting layers,
which can be an advantage of this scheme compared to other
surface-sensitive spectroscopic methods. Recently, however, it has
been suggested that the experimental observations of the IJJs could
be affected by the self-heating generated in a high-bias region,
which was caused by the poor thermal conductivity of
Bi$_2$Sr$_2$CaCu$_2$O$_8$ (Bi-2212) and other materials involved in
the tunneling measurements.\cite{Yurgens}

The zero-bias tunneling process in the $c$ axis is very sensitive
to the electronic DOS at the normal-state Fermi surface. In
particular, the zero-bias tunneling resistance $R_c$ in the
Bi-2212 is weighted by the tunneling of quasiparticles in the
antinodal region of the Fermi surface.\cite{Hop} Thus, $R_c$ is
expected to increase rapidly as the PG opens and the corresponding
DOS is partially depleted at the Fermi surface. In this point of
view, the onset temperature of the PG opening can be defined as
the characteristic temperature $T^*_{dev}$, at which $R_c$
deviates from the $T$-linear temperature dependence in $R_c$ vs
$T$ curves.\cite{RT}

Recently, Kawakami {\it et al.},\cite{RT} based on the temperature
dependence of $R_c$, have shown that both the electron- and the
hole-doped cuprates have common spin-singlet correlations in
forming the PG, both of which thus close in high magnetic fields.
The closing fields of the PG and the SG, however, show much
different temperature dependencies from each other, which
indicates that the two gaps are of separate origins. Difference in
origins of the two gaps is in line with the coexistence of the
superconducting state and the PG state observed by the interlayer
tunneling spectroscopy (ITS) in hole-doped Bi-2212 IJJs, which is
represented by the sharp peak and the broad hump structure below
$T_c$.\cite{ITS} Relating the hump structure in the high-energy
windows of the ITS to the formation of the PG was controversial up
to the present, however, again because of possible self-heating in
a high-bias region,\cite{Fischer} although there have been many
efforts to reduce the self-heating effect in the ITS, by adopting
schemes such as reducing the junction area, reducing the number of
stacked junctions, and adopting pulsed
biasing.\cite{ITS,Krasnov4,Zhu}

In this study, for an overdoped Bi-2212 sample fabricated on an
as-grown single crystal, we discriminate the PG onset temperature,
defined by the appearance of the hump structure ($T^*_{hump}$) in
the ITS while lowering temperature, from that obtained by the
$R_c$ vs $T$ behavior ($T^*_{dev}$). To obtain the interlayer
tunneling characteristics that are essentially free from
self-heating artifact, we adopted the recently developed technique
of heating-compensated ITS, where a large portion of the
bias-induced self-heating was removed.\cite{Bae2} In contrast to
$T^*_{dev}$$\sim$ 190 K, the hump structure persisted up to
temperatures much higher than commonly perceived. We then
numerically illustrate that, with significant self-heating,
$T^*_{hump}$ would reduce down to $T^*_{dev}$, which confirms the
previous reports by others in the heating-dominated
case.\cite{Yurgens} We thus suggest that the $T^*_{hump}$
represents the genuine onset of the PG. As in earlier
works\cite{Krasnov}, with decreasing temperature below $T_c$, the
hump voltages in our heating-free ITS increases, along with the
increase of the superconducting gap size. It turns out that this
unusual temperature variation of the hump voltage below $T_c$
results from the combined tunneling contribution of the
quasiprticles associated with both the SG and the PG in the
electronic state. Since the antinodal tunneling is weighted, the
behavior of this PG obtained in the ITS of our overdoped Bi-2212
should be related to the electronic structure of the antinodal
region in the first brillouin zone of the CuO$_2$ layers.

\section{Experiment}

\begin{figure}[t]
\begin{center}
\leavevmode
\includegraphics[width=0.8\linewidth]{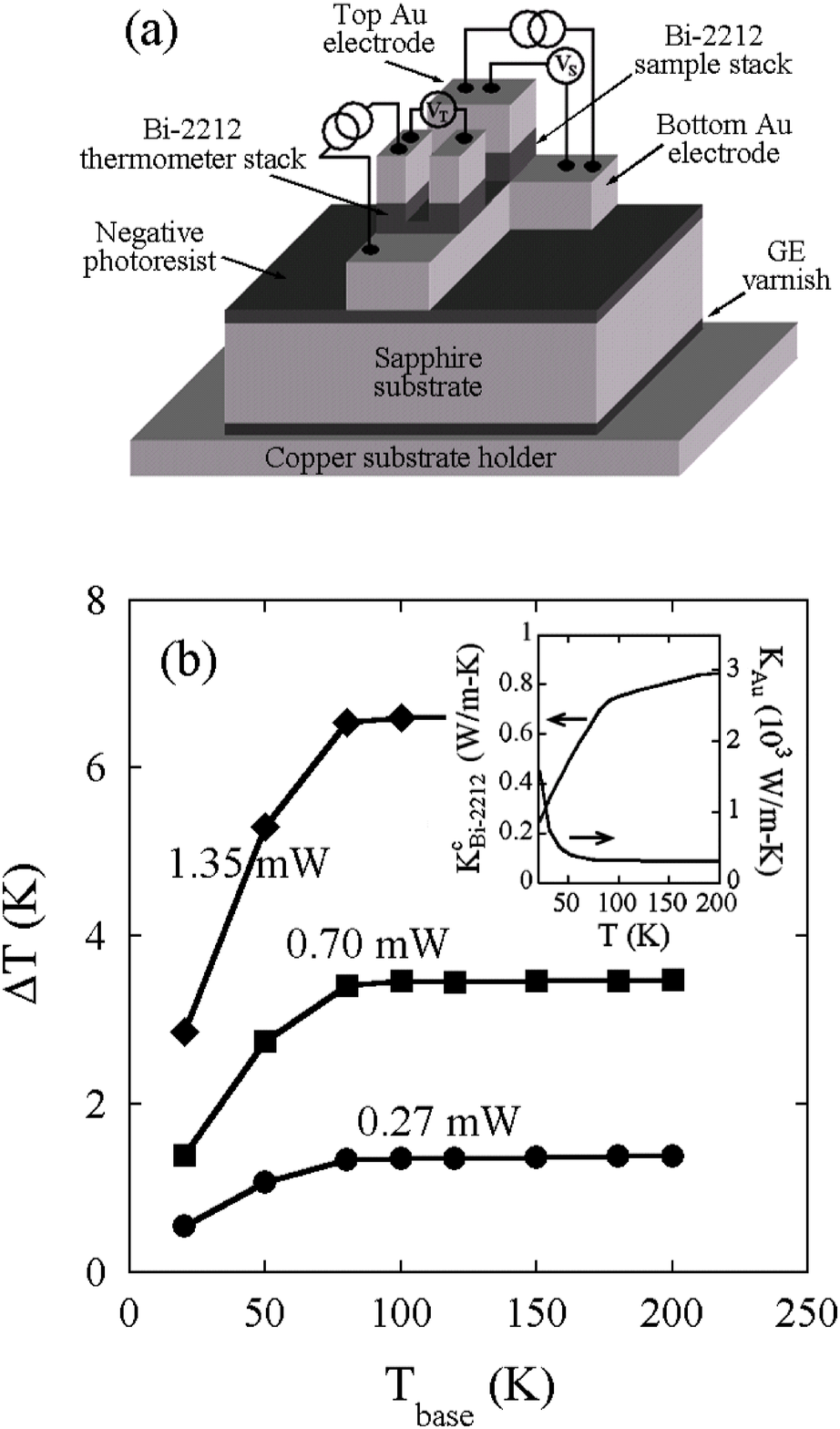}
\caption{(a) Schematic sample configuration adopted to estimate the
temperature profiles in the sample. The thermometer stack is assumed
to be 0.5 $\mu$m apart laterally from the sample stack. (b) The
temperature discrepancies between the sample stack and thermometer
stacks as a function of base temperature of the substrate holding
copper block for various heating powers dissipated in the sample
stack. Inset: the thermal conductivities of Bi-2212 along the c-axis
direction and Au electrode material as a function of temperature.}
\end{center}
\end{figure}

Fig. 1(a) illustrates the schematic configuration of the sample. We
fabricated an overdoped sample stack of Bi-2212 IJJs with the
lateral size of 3$\times$3 $\mu$m$^2$, sandwiched between two
thin-film (top; 400 nm thick, bottom; 100 nm thick) Au electrodes.
This structure, where the pedestal stack (large stack of IJJs
outside of but coupled to the stack of IJJs of interest) in the
usual mesa structure was eliminated, gave more uniform tunneling
current distribution.\cite{Bae} The number of junctions contained in
the stack, $N$, was 19. The hole concentration, $p$= 0.19, was
determined using the $c$-axis superconducting transition temperature
$T_c$=88.3 K by the empirical relation of
$T_c$=95[1-82.6($p$-0.16)$^2$] and the resistance ratio of $R_c$
between $T_c$ and $T$=300 K.\cite{Watanabe} A thermometer stack of
IJJs, with the lateral dimension of 3$\times$2 $\mu$m$^2$, was
arranged less than 1 $\mu$m from the sample stack through a
100-nm-thick bottom Au electrode.

For the ITS the whole probe with a sample inside the vacuum can
was cooled down to the liquid-helium bath temperature. Prior to
the ITS measurements in a finite bias the sample was set at a
higher working temperature by using a resistive heater coil wound
on the substrate-holding copper block. The temperature variation
of the sample stack during ITS measurements was monitored {\it
in-situ} by the change in the tunneling resistance of the
thermometer stack taken in a constant bias current\cite{Bae2} of
$I_{th}=120$ $\mu$A. To maintain the thermometer stack at a given
set temperature during ITS measurements we compensated the
bias-induced heating by lowering the current level to the heater
coil wound on the copper substrate holder block. We repeated this
heating-compensation scheme by using a computer-aided
proportional-integral-derivative control of the thermometer stack
incorporated with adjusting the heating-current level. By adopting
this technique we were able to maintain the temperature of the
thermometer stack within about 0.2 K in the whole bias range of
the ITS.\cite{low limit} However, there could still be a
temperature difference, $\Delta T$, between the sample stack and
the thermometer stack due to the finite thermal conductance of the
bottom Au electrode, through which the heat generated in the
sample stack flowed to the thermometer stack, and due to the
thermal leakage of heating to the surroundings.\cite{Zavaritsky}

We numerically estimated the temperature difference $\Delta T$
between the sample stack and the thermometer stack during ITS
measurements. The COMSOL Multiphysics Program was used to
calculate the temperature profile in the sample.\cite{COMSOL} In
the estimation, we referred to the geometry and arrangement of the
sample used in the measurement [Fig. 1(a)]. Namely, the common
bottom Au electrode was attached to the sapphire substrate (0.4 mm
thick and 5$\times$5 mm$^2$ in the lateral size) using negative
photoresist, which in turn was fixed on the heater-coil-wound
copper block using GE varnish (assumed to be about 1 $\mu$m
thick). The top Au electrodes in the sample and thermometer stacks
were extended by up to 200 $\mu$m by two (300-nm-thick and
3-$\mu$m-wide) Au stripes (per each stack) deposited on the
substrate that was precoated with an about 1-$\mu$m-thick negative
photoresist layer. The end of each Au extension was then connected
to a gauge-$\sharp$40 copper wire that was thermally anchored at
the base copper-block temperature. The bottom common electrodes
were also extended in the same manner by three Au stripes; two in
the sample-stack side and one in the thermometer-stack side.

The inset of Fig. 1(b) shows the temperature dependence of the
thermal conductivities of Au and Bi-2212 along the $c$
axis,\cite{Au,Crommie-Bi2212} which played a crucial role in the
heat flow through the sample. The in-plane thermal conductivity of
Bi-2212 was assumed to be ten times higher than the $c$-axis
thermal conductivity.\cite{Krasnov3} The thermal conductivities
for sapphire, negative photoresist, and GE varnish were assumed to
be insensitive to the temperature variation and set to be 40, 0.2,
and 0.2 W/m-K, respectively. The heat generated at the sample
stack was dissipated through both the top and bottom thermal
channels, while the temperature of the thermometer stack was
maintained at a fixed temperature during the heating-compensated
ITS. Fig. 1(b) shows the discrepancy of temperature between the
the sample stack and the thermometer stack as a function of the
copper-block base temperature, corresponding to the heating power
at the sample stack of 0.27, 0.70, and 1.35 mW. One notices that
$\Delta T$ is governed by the temperature dependence of the
thermal conductivity of the bottom Au electrode: $\Delta T$
increases in the temperature range of 20-80 K because of reduction
of the Au thermal conductivity until it saturates at temperatures
above 100 K. The heating power of 0.70 mW corresponds to the bias
voltage that is high enough to observe the hump structure in the
differential conductance in the inset of Fig. 2 ($V$=525 mV) and
in Figs. 4 and 5(a) ($v$=$V$/19=27.6 mV). Then the estimation
indicates that the heating-compensated thermometry adopted in this
study allowed accuracy of the thermometry within 3.5 K for the
heating power of 0.70 mW at any base temperatures. This is in
remarkable contrast to the discrepancy of several tens of
degrees,\cite{Yurgens} usually encountered without the heating
compensation incorporated with the {\it in-situ} thermometry.
Thus, the hump structure obtained from this heating-compensated
ITS can be regarded to be almost heating-free.

\section{Results and discussion}

\begin{figure}[b]
\begin{center}
\leavevmode
\includegraphics[width=0.8\linewidth]{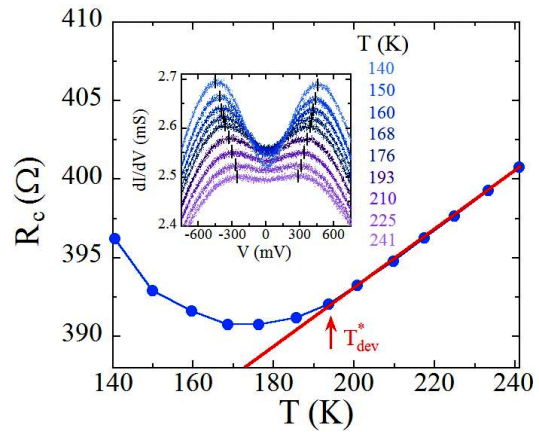}
\caption{(color online) Temperature dependence of
$R_c$[=$dI/dV|^{-1}_{V=0}$] curves of the overdoped sample. The
straight line in the figure is a guide to the eye, showing linear
$R_c$ above a characteristic temperature $T^*_{dev}$. $R_c$ may
increase rapidly as the pseudogap opens and the corresponding
electron density of states is partially depleted at the Fermi
surface. In this point of view, $T^*_{dev}$ is often defined as the
onset temperature of pseudogap opening.\cite{RT} The inset:
interlayer tunneling spectra $dI/dV$, showing hump structures in the
normal state above $T_c$.}
\end{center}
\end{figure}

Fig. 2 shows the $R_c$ vs $T$ curve in the normal state of the
sample, which was obtained by the inverse of zero-bias tunneling
conductance $dI/dV|_{V=0}$ of the electronic excitation spectra in
the inset of Fig. 2. This curve suggests the PG onset temperature
to be $T^*_{dev}\sim$190 K. The normal state of our overdoped
sample in the inset of Fig. 2 shows a distinct zero-bias depletion
of electronic excitation spectra with the PG size for each $T$
denoted by a pair of vertical segments. The onset temperature of
the PG opening can also be determined by the appearance of the
hump structure ($T^*_{hump}$) in the tunneling $dI/dV$ spectra,
which can be obtained either by the scanning tunneling
spectroscopy\cite{Fischer} (STS) or by the ITS. In contrast to a
naive expectation and the existing observations,\cite{RT,ITS}
however, the depletion of the DOS near zero bias is evident even
far above $T^*_{dev}$ in $R_c$ vs $T$ curves. The spectral
depletion around zero bias persists up to the maximum temperature
examined, {\it i.e.}, 241 K. We observed similar behavior in
another overdoped sample. But, the deviation from the $T$-linear
behavior in $R_c$ vs $T$ is supposed to become evident only when
the DOS is sufficiently depleted at the characteristic temperature
$T^*_{dev}$ further below the onset temperature of the hump
structure opening. This indicates that $T^*_{hump}$ better
represents the onset of PG opening than $T^*_{dev}$. On the other
hand, outside the gapped region (${\it i.e.}, $ $|V|>$450 mV) in
the inset of Fig. 2, $R_c$ monotonically increases with increasing
temperature over the whole temperature range examined, which
represents a metallic behavior. This PG onset temperature defined
by $T^*_{hump}$ at least in the overdoped regime is in clear
contrast to that determined by the angle-resolved photoemission
spectroscopy (ARPES) and by the STS,\cite{Timusk} where $T^*$,
representing the PG in the one-gap picture, disappears or merges
into the bell-shaped $T_c$ curve near the optimal doping point in
the temperature-vs-doping-level phase diagram.

On the other hand, $T^*_{hump}$ becomes comparable to  $T^*_{dev}$
in the presence of significant self-heating. Based on the
heating-compensated $dI/dV$ curves in the normal state of the
inset of Fig. 2, we simulated the $dI/dV$ curves of a stack under
the influence of the serious self-heating. Fig. 3(a) displays
$R_c$-vs-$T$ curves obtained from the $dI/dV$($V$) spectra given
in the inset of Fig. 2 for varying biases from 0 to 360 mV (or
from 0 to 18.9 mV per junction). With increasing the bias voltage,
the up-turn deviation from the $T$-linear $R_c$-vs-$T$ behavior
gradually disappears. The sample temperature increases by a bias
power defined at a fixed voltage with a given heating ratio
[K/mW]. The heating ratio in a stack of IJJs is determined by the
junction area and the number of junctions. The new value of
$dI/dV$ at an increased temperature due to self-heating for a
finite bias voltage was traced through the $R_c$-vs-$T$ curves of
the corresponding voltage in Fig. 3(a). Since current-voltage
({\it I-V}) characteristics had almost the same curvatures in the
temperature range under investigation we assumed that the power at
a fixed voltage remained almost the same in the temperature range.

\begin{figure}[t]
\begin{center}
\leavevmode
\includegraphics[width=0.8\linewidth]{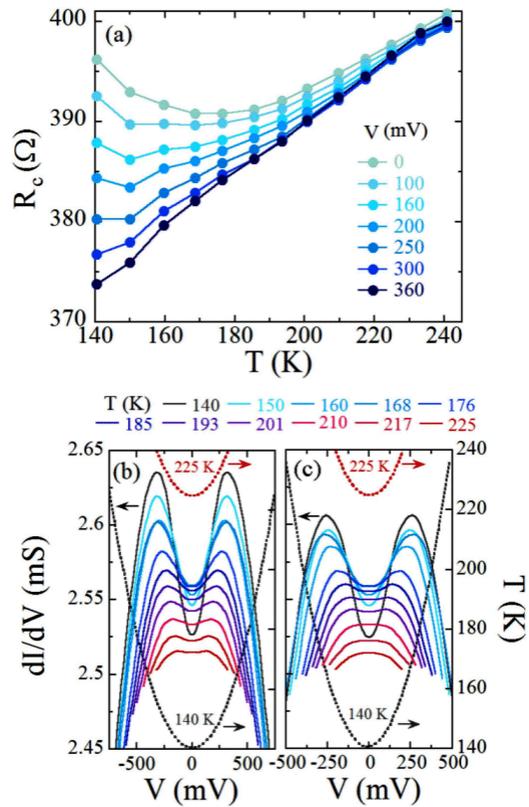}
\caption{(color online) (a) Temperature dependence of $R_c$ with
increasing bias voltages. (b) and (c) Temperature dependence of
$dI/dV$ curves estimated based on the curves in (a) for finite
self-heating with the heating ratio of 60 and 150K/mW, respectively.
The dashed curves in (b) and (c) illustrate the temperature
variation from the base temperature of 140K and 225 K, respectively,
by the bias-induced self-heating for the two values of the heating
ratio.}
\end{center}
\end{figure}

Figs. 3(b) and 3(c) show the calculated $dI/dV$ curves, which would
be affected by self-heating with the heating ratio of 60 K/mW and
150 K/mW, respectively. Figures also show the effective temperature
of the sample as a function of the bias voltage for two base
temperatures, $T_b$=140 K and 225 K. Since the hump structures in
these figures are weakened by the self-heating the voltage positions
and the heights of the humps get smaller than the ones shown in the
inset of Fig. 2. In Fig. 3(c), a higher heating ratio than that in
Fig. 3(b) makes the hump structure disappear at a lower temperature
around $T$$\sim$210 K, which is close to $T^*_{dev}$. This
calculation clearly shows that the disappearance of the hump
structure near $T^*_{dev}$ is indeed due to
self-heating.\cite{Yurgens} Our nonlinear $dI/dV$ curves in the
normal state are in contrast to the flat $dI/dV$ behavior modeled
for the normal state in Ref. 24. The hump structures for
temperatures above $\sim$170 K as in the inset of Fig. 2, which lead
to the local minimum of $R_c$ in the main panel, cannot be
explained, either, in terms of the self-heating model with a flat
$dI/dV$ behavior. Therefore, the hump structure previously reported
in the ITS\cite{Yurgens,Krasnov} should not have been solely from
the self-heating effect but from the intrinsic depletion of the
zero-bias electronic spectral weight, presumably affected by the
self-heating. This PG behavior with high onset temperature, $T^*$,
in the overdoped regime, revealed by our heating-compensated ITS,
has characteristics similar to that observed in the angle-integrated
photoemission spectroscopy and in the electronic magnetic
susceptibility $\chi(T)$.\cite{Nakano} This peculiar PG behavior
displays a clear peak-dip-hump structure below but near $T_c$, where
the peak pertains to the superconducting coherence.

\begin{figure}[t]
\begin{center}
\leavevmode
\includegraphics[width=0.8\linewidth]{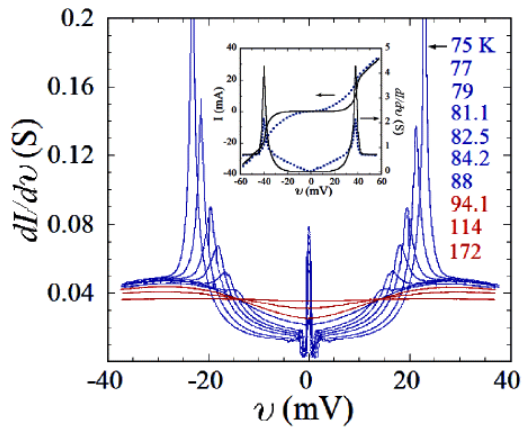}
\caption{(color online) The heat-compensated interlayer tunneling
spectra $dI/dv$ for our overdoped sample as a function of bias
voltage per junction ($v$) at various $T$. Inset: {\it I-v} and
$dI/dv(v)$ curves calculated using Eq. (1) for $T_{\phi}$=1 (dotted
curves) and $T_{\phi}$=$t_{\bot}\mbox{cos}^2 2\phi$ (solid curves)
at $T$=4.2 K.}
\end{center}
\end{figure}

A series of the overall feature of heating-compensated interlayer
tunneling spectra $dI/dv(v)$ of our overdoped sample, for varying
$T$, is displayed in Fig. 4, where the voltage is normalized by
the number of junctions as $v$=$V/N$. In the normal state above
$T_c$ the low-bias DOS is smoothly depleted, revealing the PG. At
a $T$ below $T_c$ a sharper peak (the coherence peak) develops
inside the PG, constituting the peak-dip-hump structure. Further
lowering $T$, the fast sharpening coherence peak with the growing
SG size overwhelms the spectrum, leaving only the coherence peak
apparent. The PG with the hump becomes more conspicuous in
underdoped samples (not shown). Below $T_c$, the tunneling spectra
show the more U-shaped DOS in the subgap region than the one
observed previously.\cite{ITS} The fluctuating conductance at zero
bias sufficiently below $T_c$ in Fig. 4 was caused by the
Josephson pair tunneling.

\begin{figure}[b]
\begin{center}
\leavevmode
\includegraphics[width=0.8\linewidth]{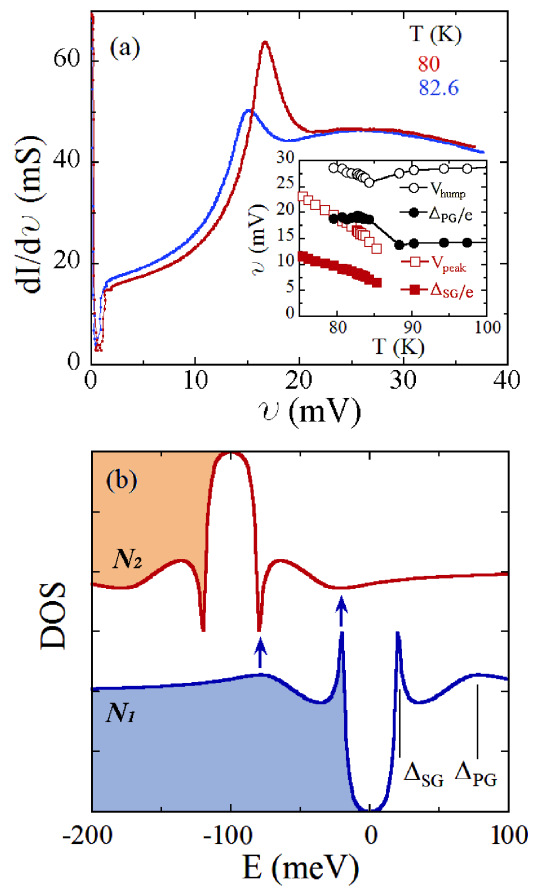}
\caption{(color online) (a) $dI/dv$ curves for $T$=80 and 82.6 K in
our overdoped sample. (b) the shape of imagined electron density of
states of two identical superconducting electrodes of a junction,
which contain both the superconducting coherence peak and the PG
hump, with assumed gap sizes of $\Delta_{SG}$=20 meV and
$\Delta_{PG}$=80 meV. One electrode is assumed to be biased with
$V$=100 mV. Inset of (a): temperature dependence of voltage
positions of the superconducting coherence peak ($V_{peak}$) and the
PG hump ($V_{hump}$), along with the corresponding SG energy
($\Delta_{SG}$) and PG energy ($\Delta_{PG}$).}
\end{center}
\end{figure}

In the following we discuss features of the superconducting gaps
and pseudogaps and the interrelation between them one can observe
in or deduce from the heating-free ITS results. Fig. 5(a) shows
$dI/dv(v)$ curves at $T$=80 K and 82.6 K below $T_c$ of our
overdoped sample. The electronic spectra near and below $T_c$ show
a peak-dip-hump structure. As the coherence peak in the ITS starts
to develop below $T_c$ the height and voltage position of the peak
increases with decreasing temperature as illustrated in Figs. 4
and 5(a). This represents the SG edge. The hump structure in the
normal state in the inset of Fig. 2 is connected to that below
$T_c$, which thus represents the PG state. This observation is
consistent with the recently reported Raman response functions,
which reveal a change of spectral weight from lower energies to
higher ones when making a transition from the normal state into
the superconducting state at the optimal doping and slightly
overdoped levels. In particular, the changes of the Raman
responses are more enhanced in the antinodal regime than in the
nodal regime.\cite{Tacon} The coexistence of the superconducting
and PG states below $T_c$ reflects that the two gaps are of
different origins.\cite{ITS} This is in contradiction to the
characteristics of the one-gap concept observed in ARPES and STS,
where the PG is smoothly connected to the SG near $T_c$ in the
varied doping ranges.\cite{Timusk,Fischer}

The hump structure in the peak-dip-hump excitation spectral
distribution provides valuable information on the interrelation
between the PG and superconducting states. The inset of Fig. 5(a)
shows the temperature dependence of the superconducting coherence
peak voltage ($V_{peak}$) and the PG hump voltage ($V_{hump}$)
near $T_c$. Since the tunneling occurs between two neighboring
superconducting layers, the SG energy $\Delta_{SG}$ should be one
half of $eV_{peak}$. In fact, the value of $V_{peak}$ is seen to
reduce rapidly along with the SG as the temperature approaches to
$T_c$(=88 K) from below. However, it turns out that the PG energy
$\Delta_{PG}$ determined in relation with $eV_{hump}$, for the
superconductor-insulator-superconductor (SIS) tunneling near
$T_c$,\cite{Mourachkine} should be defined in a somewhat different
way. For temperatures in the range of $T$$>$$T_c$, where
$\Delta_{SG}$ vanishes completely, $\Delta_{PG}$ is supposed to be
simply $eV_{hump}$/2. But for $T$$<$$T_c$ the value of $V_{hump}$
is affected by opening of the SG as well as the PG.

The differential conductance as a function of the voltage bias in
the SIS tunneling junction is given by\cite{Tinkham}
\begin{eqnarray}
\frac{dI}{dV}&\propto&\frac{d}{dV}\int_0^{2\pi}d\phi
|T_{\phi}|^2\int N_1(E,\phi)N_2(E+eV,\phi)\nonumber \\
& & \times [f(E,T)-f(E+eV,T)]dE,
\end{eqnarray}
where $N_1$ and $N_2$ are the electron DOS of two identical
superconducting layers with the $d_{x^2-y^2}$ symmetry, $T_{\phi}$
is the tunneling matrix and $\phi$ [=tan$^{-1}(k_y/k_x)$] is the
azimuthal tunneling angle. The angle-integrated electronic DOS,
$N_1(E)=\int_0^{2\pi}d\phi N_1(E,\phi)$ for an assumed
$N_1(E,\phi)$ in the presence of both the superconducting and the
PG states, is illustrated in Fig. 5(b). Here, two different kinds
of quasiparticles dressed by the superconducting state and the PG
state are assumed to be accumulated near $\Delta_{SG}$=20 meV and
$\Delta_{PG}$=80 meV, respectively. Quasiparticles fill all states
below zero bias. If the voltage bias, $V$, is applied to the
counter-electrode superconductor $N_2$ is shifted by $eV$ along
the energy axis. In this process, quasiparticles with an energy
$E$ in the occupied region of $N_1$ tunnel to the unoccupied
states of $N_2$ at the same energy. This tunneling weight
determines the conductance at a given voltage bias in the
measurement. Especially, superconducting coherence peak at
$E$=$\Delta_{SG}$ of $N_1$ dominantly determines the conductance
shape as a function of voltage if the peak is sufficiently higher
than the hump in the electronic DOS.

The angle-integrated DOS of the counterelectode, $N_2(E)$, shown
in Fig. 5(b) is the case where the bias voltage $V$=100 mV
corresponds to ($\Delta_{SG}$+$\Delta_{PG}$)/e. Here, the
superconducting coherence peak and the PG hump filled with the
quasiparticles of $N_1$ are respectively overlapped with the
vacant PG hump and the superconducting coherence peak of $N_2$. In
this condition, the conductance is enhanced at the voltage bias
corresponding to the hump voltage in the $dI/dV(V)$ curves. Thus,
below $T_c$, $eV_{hump}$=$\Delta_{SG}$+$\Delta_{PG}$ rather than
simply $2\Delta_{PG}$. This inclusion of the influence of opening
of the SG, $\Delta_{SG}$, in $V_{hump}$ \emph{explains the reason
why $V_{hump}$ increases along with $V_{peak}$ below $T_c$} as
seen in the inset of Fig. 5(a). Thus, care should be taken in
extracting gap values from $dI/dV(V)$ curves of an SIS junction.
If the height of superconducting coherence peak is not much higher
than that of the PG in the tunneling DOS close to $T_c$, the
$eV_{hump}$ turns out to be positioned between
$\Delta_{SG}+\Delta_{PG}$ and $2\Delta_{PG}$. Thus, the $V_{hump}$
can be smoothly connected near $T_c$ as in the inset of Fig. 5(a)
and also as reported previously.\cite{ITS,Krasnov} If
$\Delta_{SG}$ gets close in its value to $\Delta_{PG}$ at
sufficiently low temperatures, one cannot easily distinguish the
hump position from the coherence peak position because
$\Delta_{SG}$+$\Delta_{PG}$$\sim$$2\Delta_{SG}$. This is the
reason why no hump structure is visible below $\sim$80 K in the
inset of Fig. 5(a). \cite{Krasnov}

The inset of Fig. 5(a) also displays the temperature dependence of
the two characteristic gaps extracted using the above analysis.
With decreasing temperature near and below $T_c$, the PG increases
abruptly. This unusual behavior arises because the apparent hump
structure in the tunneling spectra of a junction largely depends
on which spectral weight, the SG or the PG, of an electrode is
coupled to the PG spectral weight of the counterelectrode; (i) for
$T$$>$$T_c$, the PG DOS in one electrode is detected by the broad
pseudogap DOS in the counterelectrode, which makes $\Delta_{PG}$
lower than the expected position because of the broadness and (ii)
$T$$<$$T_c$, PG DOS in one electrode is detected by the sharp
superconducting peak in the opposite electrode, giving the value
of $\Delta_{PG}$ close to the expectation.

This DOS analysis indicates that the hump structure at temperature
below $T_c$ is due to tunneling of quasiparticles associated with
the SG (PG) in one electrode of a junction to a vacant
quasiparticle state associated with the PG (SG) in the counter
electrode. In this tunneling process of the quasiparticles, the
key observation is that \emph{the quasiparticle tunneling
constituting the hump structure is possible only if the SG and the
PG arise from the combined electronic state} of quasiparticles of
two different characters in the same momentum space, {\it i.e.},
the antinodal region. This picture implies that quasiparticles in
a single state distribute either in the SG spectra or in the PG
spectra, depending on external physical parameters such as
temperature, magnetic field, doping, etc. This inference is
consistent with our earlier observation that, in a several-tesla
$c$-axis magnetic field, the tunneling spectral weight in stacks
of Bi-2212 IJJs (for both overdoped and underdoped ones)
redistributes from the superconducting coherence peak to the PG
hump.\cite{pseudogap} But, the fact that the superconducting and
the PG states are based on the single electronic structure
composed of quasiparticles of two different characters is in
contradiction to the earlier tunneling measurements with claiming
that the peak-dip-hump features arise from a simple overlap of
spectral functions of antibonding and bonding states, associated
with the bilayer splitting\cite{Hoogenboom}.

The anisotropic tunneling matrix element\cite{Hop} in the
interlayer tunneling in the Bi-2212 filters out the tunneling in
the nodal region and weights the tunneling near the antinodal
points on the momentum space in the Brillouin zone. The pronounced
U-shape in the measured tunneling $dI/dV$ curves of Figs. 4 and
5(a), which is in contrast to the V-shape ones usually observed in
the STS,\cite{Hoogenboom,Fischer} is caused by this filtering. The
dotted curves in the inset of Fig. 4 show the numerically obtained
{\it I-v} ($v$ is the bias voltage per junction) and the
differential conductance curves using Eq. (1) for a
$k$-independent tunneling matrix element
$T_{\phi}$=1,\cite{Yamada} with $\Delta_{SG}$=20 meV and the
quasiparticle scattering rate $\Gamma$= 0.05 meV for an assumed
DOS, $N(E, \phi)=\mbox{Re}\{(E-i\Gamma)/[(E-i\Gamma)^2-
\Delta_{SG}^2\mbox{cos}^22\phi]^{1/2}\}$ at $T$=4.2 K. The solid
curves in the inset of Fig. 4 correspond to the anisotropic
tunneling matrix element $T_{\phi}=t_{\bot}\mbox{cos}^2 2\phi$,
which is theoretically predicted for a crystal with the tetragonal
symmetry as Bi-2212.\cite{Hop} Here, $t_{\bot}$ is the hopping
constant. This anisotropic $T_{\phi}$ reduces the low-energy
quasiparticle tunneling near the nodal points and leads to the
U-shaped tunneling conductance, while sharpening the coherence
peak \cite{Su}. Thus, the ITS in our heating-compensation scheme
mainly shows the electronic state in the vicinity of the antinodal
region and the PG formation is more closely related to the
electronic state in this region.\cite{Hashimoto} It is widely
accepted that a Fermi-surface nesting exists with a van Hove
singularity (high DOS with a flat band) near the antinodal
region,\cite{Gofron} which is related to the formation of an
antiferromagnetic order or orders like the spin-density wave and
the charge-density wave. This Fermi surface nesting may be related
to the downturn behavior of background spectra of the
ITS\cite{Krasnov} in the inset of Fig. 2. The Fermi nesting near
the antinodal region is reduced with increasing the doping in
hole-doped cuprates because of change of the Fermi surface
topology with doping.\cite{Flat} The angle-integrated
photoemission spectroscopy also showed that the binding energy of
the PG corresponding to the flat-band position of the antinodal
region, the so-called high-energy pseudogap, decreases with
increasing doping.\cite{AIPES} Thus, one can expect that the PG
energy scale and the PG onset temperature will decrease with
increasing doping if the PG in the ITS is associated with the
antinodal electronic state. Indeed, it has been reported that the
PG onset temperature $T^*_{dev}$ and the PG closing field $H_{pg}$
observed in $R_c$ vs $T$ decrease with increasing
doping.\cite{RT,RT2}

The low-energy pseudogap is believed to be a precursor of the
superconducting state, while the high-energy pseudogap (HEPG) is
inferred to be of an antiferromagnetic order or orders like the
spin-density wave and the charge-density wave. In the hole-doped
cuprates, the low-energy pseudogap tends to close near the optimal
doping but the HEPG persists even in the heavily overdoped regime.
Features of the HEPG related to the antinodal region have been
observed in various experiments such as the electronic magnetic
susceptibility,\cite{Oda} the Knight shift,\cite{Ishida} and the
angle-integrated photoemission spectroscopy.\cite{Sato} The onset
temperatures of the HEPG over the doping values in these
measurements were almost twice as high as those of the low-energy
pseudogap. Especially, the onset temperatures of the HEPG for
$p$$\sim$0.19 were 260$\sim$270 K, close to our ITS results at the
same doping value. Since the interlayer tunneling spectra mainly
reveal the electronic state in the vicinity of the antinodal
region the formation of the pseudogap is closely related to the
electronic state in this region. It has been widely accepted that
the antinodal regions in cuprates are related to the formation of
an antiferromagnetic order or orders like the spin-density wave
and the charge-density wave. It thus strongly indicates that our
hump structure is highly likely to be related to the HEPG.

\section{Conclusion}

In order to understand the nature of the PG, the interlayer
tunneling spectroscopic characteristics of high-$T_c$
superconductors were investigated, while the self-heating was
largely excluded using the heating-compensation technique
incorporated with the {\it in-situ} thermometry. Since the ITS is
sensitive to the bias-induced self-heating an extreme care should
be taken, as in this study, to keep the sample temperature
constant within a tolerance limit over the whole bias sweeping
range. But, as demonstrated by Krasnov {\it et
al.},\cite{Krasnov4, Krasnov5} most of the essential observations
on the PG feature remain valid even in the earlier ITS results.
Thus, ITS, with a precautious measure taken to eliminate the
self-heating, should provide a very useful experimental tools to
investigate the electronic excitation spectrum of highly
anisotropic materials containing naturally grown tunneling
junctions.

In this study, it is found that the genuine PG behavior in the
overdoped cuprates reveals the following characteristics. Defined
by the appearance of the hump at $T^*_{hump}$ in the differential
tunneling conductance, the PG onset temperature $T^*$ reaches up
to nearly room temperature, much higher than the estimation based
on the tunneling resistive transition, $T^*_{dev}$. With
significant self-heating, however, numerical simulation shows that
$T^*_{hump}$ returns to $T^*_{dev}$. This observation indicates
that the hump structure in the tunneling differential conductance
provides far more accurate determination of the PG onset
temperature than the tunneling resistive transition. The hump
voltage revealed in the ITS below $T_c$ is shown to follow the SG
value, which is, in fact, additional confirmation that the hump
structure in the ITS represents a genuine electronic PG state. The
hump structure in the ITS below $T_c$ is also affected by the
relative height and the voltage of the superconducting coherence
peak, which originates from the fact that tunneling quasiparticles
in an IJJ are dressed by the presence of both the SG and the PG.
Since the interlayer tunneling is sensitive to the electronic
state in antinodal region existing in a flat band the PG behavior
in ITS, coexisting with the SG, should be related to the Fermi
surface nesting induced by the van Hove
singularity.\cite{Markiewicz}

\section*{ACKNOWLEDGMENTS}

We want to acknowledge illuminating private communications with
H.-Y. Choi. One of us (H.-J. Lee) appreciates
valuable discussion and communications with V. M. Krasnov and A.
Yurgens on the thermometry in the presence of bias-induced
heating. This work was supported by the National Research
Laboratory program administrated by Korea Science and Engineering
Foundation (KOSEF). This paper was also supported by POSTECH Core
Research Program and the Korea Research Foundation Grants No.
KRF-2006-352-C0020.


\begin{thebibliography}{00}
\bibitem{Timusk} T. Timusk and B. Statt, Rep. Prog. Phys. {\bf 62}, 61 (1999) and the references cited therein.
\bibitem{Damascelli} A. Damascelli, Z. Hussain, and Z.-X. shen, Rev. Mod. Phys. {\bf 75}, 473 (2003).
\bibitem{Arc} M. R. Norman, H. Ding, M. Randeria, J. C. Campuzano, T. Yokoya,
T. Takeuchi, T. Takahashi, T. Mochiku, K. Kadowaki, P. Guptasarma, and D. G. Hinks
, Nature (London) {\bf 392}, 157 (1998); A. Kanigel, M. R. Norman,M.
Randeria, U. Chatterjee, S. Souma, A. Kaminski, H. M. Fretwell, S.
Rosenkranz, M. Shi, T. Sato, T. Takahashi, Z. Z. Li, H. Raffy, K.
Kadowaki, D. Hinks, L. Ozyuzer, and J. C. Campuzano, Nature Phys.
(London) {\bf 2}, 447 (2006).
\bibitem{Dessau} D. S. Dessau, Z.-X. Shen, D. M. King, D. S. Marshall,
L. W. Lombardo, P. H. Dickinson, A. G. Loeser, J. DiCarlo, C.-H
Park, A. Kapitulnik, and W. E. Spicer, Phys. Rev. Lett. {\bf 71},
2781 (1993).
\bibitem{Gofron} K. Gofron, J. C. Campuzano, A. A. Abrikosov, M. Lindroos, A. Bansil,
H. Ding, D. Koelling, and B. Dabrowski, Phys. Rev. Lett. {\bf 73}, 3302 (1994).
\bibitem{Gabovich} A. M. Gabovich, A. I. Voitenko, J. F. Annett, and M. Ausloos, Supercond. Sci. Technol {\bf 14}, R1 (2001).
\bibitem{Varma} C. M. Varma and L. Zhu, Phys. Rev. Lett. {\bf 98}, 177004 (2007);
A. P. Kampf and J. R. Schrieffer, Phys. Rev. B {\bf 42}, 7967 (1990);
X.-G. Wen and P. A. Lee, Phys. Rev. Lett. {\bf 76}, 503 (1996);
I. S. Elfimov, G. A. Sawatzky and A. Damascelli, arXiv:0706.4276v1;
N. Doiron-Leyraud, C. Proust, D. LeBoeuf, J. Levallois, J.-B. Bonnemaison,
R. Liang, D. A. Bonn, W. N. Hardy and L. Tailler, Nature (London) {\bf 447}, 565 (2007).
\bibitem{Tacon} M. Le Tacon, A. Sacuto, A. Geoges, G. Kotliar, Y. Gallais, D. Colson and A. Forget, Nat. Phys. {\bf 2}, 537 (2006);
W. Guyard, M. Le Tacon, M. Cazayous, A. Sacuto, A. Geoges, G. Kotliar, D. Colson and A. Forget, arXiv:0708.3731v1;
S. H$\ddot{u}$fner, M. A. Hossain, Damascelli, and G. A. Sawatzky, arXiv:0706.4282v1.
\bibitem{Tanaka} K. Tanaka, W. S. Lee, D. H. Lu, A. Fujimori, T. Fujii, Risdiana, I. Terasaki,
D. J. Scalapino, T. P. Devereaux, Z. Hussain, Z.-X. Shen,  Science {\bf 314}, 1911 (2006);
T. Valla, A. V. Fedorov, Jinho Lee, J. C. Davis, G. D. Gu,  Science {\bf 314}, 1914 (2006).
\bibitem{Fischer} $\O$. Fischer, M. Kugler, I. Maggio-Aprile, Ch. Berthod, and Ch. Renner, Rev. Mod. Phys. {\bf 79}, 353 (2007).
\bibitem{Kleiner} R. Kleiner, F. Steinmeyer, G. Kunkel, and P. M\"{u}ller, Phys. Rev. Lett. {\bf 68}, 2394 (1992).
\bibitem{ITS} M. Suzuki, T. Watanabe, and A. Matsuda, Phys. Rev. Lett. {\bf 82}, 5361 (1999);
              V. M. Krasnov, A. Yurgens, D. Winkler, P. Delsing, and T. Claeson, Phys. Rev. Lett. {\bf 84}, 5860 (2000).
\bibitem{Yurgens} A. Yurgens, D. Winkler, T. Claeson, S. Ono, and Y. Ando, Phys. Rev. Lett. {\bf 90}, 147005
                 (2003); \emph{ibid}, Phys. Rev. Lett. {\bf 92}, 259702 (2004).
\bibitem{Hop} T. Xiang and J. M. Wheatley, Phys. Rev. Lett. {\bf 77}, 4632 (1996);
T. Valla, A. V. Fedorov, P. D. Johnson, Q. Li, G. D. Gu, and N.
Koshizuka, Phys. Rev. Lett. {\bf 85}, 828 (2000); M. Eschrig and
M. R. Norman, Phys. Rev. Lett. {\bf 85}, 3261 (2000).
\bibitem{RT} T. Shibauchi, L. Krusin-Elbaum, Ming Li, M. P. Maley, and P. H. Kes, Phys. Rev. Lett. {\bf 86}, 5763
            (2001); T. Kawakami, T. Shibauchi, Y. Terao, M. Suzuki, and L. Krusin-Elbaum, Phys. Rev. Lett. {\bf 95}, 017001 (2005).
\bibitem{Krasnov4} V. M. Krasnov, M. Sandberg, and I. Zogaj, Phys. Rev. Lett. {\bf 94}, 077003 (2005).
\bibitem{Zhu} X. B. Zhu, Y. F. Wei, S. P. Zhao, G. H. Chen, H. F. Yang, A. Z. Jin, and C. Z. Gu, Phys. Rev. B {\bf 73}, 224501 (2006).
The Coulomb blockade of pair tunneling may take place for a
junction size below 1 $\mu$m$^2$, which will significantly distort
the tunneling characteristics.\cite{Latyshev} Thus, merely
reducing the junction size would not be the ultimate solution of
removing the self-heating.
\bibitem{Latyshev} Yu. I. Latyshev, S.-J. Kim, V. N. Pavlenko, T. Yamashita, L. N. Bulaevskii,
Physica C {\bf 362}, 156 (2001).
\bibitem{Bae2} M.-H. Bae, J.-H. Choi, and H.-J. Lee, Appl. Phys. Lett. {\bf 86}, 232502 (2005).
\bibitem{Krasnov} V. M. Krasnov, Phys. Rev. B {\bf 65}, 140504(R)(2002).
\bibitem{Bae} M.-H. Bae, H.-J. Lee, J. Kim, and K.-T. Kim , Appl. Phys. Lett. {\bf 83}, 2187 (2003).
\bibitem{Watanabe} T. Watanabe, T. Fujii, and A. Matsuda, Phys. Rev. Lett. {\bf 79}, 2113
(1997). For the same $T_c$, an overdoped single crystal gives a
smaller resistance ratio of $R_c (T_c)$/$R_c (300$ K).
\bibitem{low limit} A limit exists in the lowest possible operation set temperature of the sample
stack above the bath temperature of 4.2 K, because in our
heat-compensation scheme the extra self-heating by the bias should
be subtracted by reducing the heater current. With lowering the
sample operation temperature one encounters a situation where the
heater current should be completely reduced, which corresponds to
the lowest limit of the sample operation temperature.
\bibitem{Zavaritsky} V. N. Zavaritsky, Phys. Rev. Lett. {\bf 92}, 259701
(2004); V. N. Zavaritsky, Phys. Rev. B {\bf 72}, 094503 (2005).
\bibitem{COMSOL} COMSOL multiphysics, $\copyright$COMSOL AB.
\bibitem{Au} G. K. White, Proc. Phys. Soc. A {\bf 66}, 559 (1953).
\bibitem{Crommie-Bi2212} M. F. Crommie and A. Zettl, Phys. Rev. B {\bf 43}, 408 (1991).
\bibitem{Krasnov3} V. M. Krasnov, A. Yurgens, D. Winkler, and P. Delsing, J. Appl. Phys {\bf 89}, 5578 (2001).
\bibitem{Nakano} T. Nakano, M. Oda, C. Manabe, N. Momono, Y. Miura, and M. Ido, Phys. Rev. B {\bf 49}, 16000 (1994); M. Oda, K. Hoya, R.
               Kubota, C. Manabe, N. Momono, T. Nakano, and M. Ido, Physica C {\bf 281}, 135 (1997); T. Sato, Y. Naitoh, T. Kamiyama, T.
               Takahashi, T. Yokoya, K. Yamada, Y. Endoh, and K. Kadowaki, Physica C {\bf 341-348}, 815 (2000).
\bibitem{Mourachkine} A. Mourachkine, Mod. Phys. Lett. B {\bf 19}, 743 (2005).
\bibitem{Tinkham} M. Tinkham, {\it Introduction to Superconductor}, 2nd ed. (McGraw-Hill, New York, 1996).
\bibitem{pseudogap} M.-H. Bae, J.-H. Choi, H.-J. Lee, and K.-S.
Park, cond-mat/0512664v1; M.-H. Bae, J.-H. Choi, H.-J. Lee and
K.-S. Park, J. of Kor. Phys. Soc. {\bf 48}, 1017 (2006).
\bibitem{Hoogenboom} B. W. Hoogenboom, C. Berthod, M. Peter, $\O$. Fischer, and A. A. Kordyuk, Phys. Rev. B {\bf 67}, 224502 (2003).
\bibitem{Yamada} Y. Yamada and M. Suzuki, Phys. Rev. B {\bf 66}, 132507 (2002).
\bibitem{Su} Y. H. Su, H. G. Luo, and T. Xiang, Phys. Rev. B 73, 134510 (2006).
\bibitem{Hashimoto} M. Hashimoto, T. Yoshida, K. Tanaka, A. Fujimori, M. Okusawa, S. Wakimoto, K. Yamada, T. Kakeshita, H. Eisaki, and S. Uchida,
Phys. Rev. B {\bf 75}, 140503(R)(2007).
\bibitem{Flat} A. Ino, C. Kim, M. Nakamura, T. Yoshida, T. Mizokawa, A. Fujimori, Z.-X. Shen, T. Kakeshita, H. Eisaki, and S. Uchida,
Phys. Rev. B {\bf 65}, 094504 (2002);  T. Yoshida, X. J. Zhou, D. H.
Lu, Seiki Komiya, Yoichi Ando, H. Eisaki, T. Kakeshita, S. Uchida,
Z. Hussain, Z.-X. Shen, and A. Fujimori, cond-mat/0610759 (2006)
\bibitem{AIPES} A. Ino, T. Mizokawa, K. Kobayashi, A. Fujimori, T. Sasagawa, T. Kimura, K. Kishio, K. Tamasaku, H. Eisaki, and S. Uchida,
Phys. Rev. Lett. {\bf 81}, 2124 (1998); T. Sato, T. Yokoya, Y. Naitoh, T. Takahashi, K. Yamada,
                and Y. Endoh, Phys. Rev. Lett. {\bf 83}, 2254
                (1999). Similar to ITS, averaging both in spatial
                and momentum spaces is taken for AIPES.
\bibitem{RT2} M. Suzuki and T. Watanabe, Phys. Rev. Lett. {\bf 85}, 4787 (2000); S. I. Vedeneev and D. K. Maude, Phys. Rev. B {\bf 70}, 184524
             (2004).
\bibitem{Oda} M. Oda, K. Hoya, R. Kubota, C. Manabe, N. Momono, T. Nakano, and M. Ido, Physica C {\bf 281}, 135 (1997).
\bibitem{Ishida} K. Ishida, K. Yoshida, T. Mito, Y. Tokunaga, Y. Kitaoka, K. Asayama, A. Nakayama, J. Shimoyama, and K. Kishio, Phys.
Rev. B {\bf 58}, R5960 (1998).
\bibitem{Sato} T. Sato, Y. Naitoh, T. Kamiyama, T. Takahashi, T. Yokoya, K. Yamada, Y. Endoh, and K. Kadowaki, Physica C {\bf 341-348}, 815 (2000).
\bibitem{Krasnov5} V. M. Krasnov, Phys. Rev. B {\bf 75}, 146501 (2007).
\bibitem{Markiewicz} R. S. Markiewicz, J. Phys. Chem. Solids {\bf 58}, 1179 (1997).
\end{thebibliography}
\end{document}